# Source Behavior for ATM ABR Traffic Management: An Explanation [1]

Raj Jain, Shivkumar Kalyanaraman, Sonia Fahmy and Rohit Goyal
Department of Computer and Information Science
The Ohio State University
Columbus, OH 43210-1277
E-mail: {*jain,shivkuma,fahmy,goyal*}@cis.ohio-state.edu

and Seong-Cheol Kim
Samsung Electronics Co. Ltd.
Chung-Ang Newspaper Bldg.
8-2, Karak-Dong, Songpa-Ku
Seoul, Korea 138-160
Email: kimsc@metro.telecom.samsung.co.kr

## Abstract

The Available Bit Rate (ABR) service has been developed to support data applications over Asynchronous Transfer Mode (ATM) networks. The network continuously monitors its traffic and provides feedback to the source end systems. This paper explains the rules that the sources have to follow to achieve a fair and efficient allocation of network resources.

# 1 Introduction

The traffic management specification TM4.0 developed by the ATM Forum is a great technological development[1, 5]. From May 1993 to April 1996, it took approximately 3 years to develop the specification. Each bimonthly meeting of the working group was attended by over 150 persons. The effort has resulted in significant advances in effective traffic management in networks covering large distances, with a large number of flows, and having a variety of quality of service (QoS) requirements.

A key distinguishing feature of ATM networks as compared to current packet networks is that they offer multiple quality of services (QoS). TM4.0 specifies five classes of services: constant bit rate (CBR), real-time variable bit rate (rt-VBR), non-real time variable bit rate (nrt-VBR), available bit rate (ABR), and unspecified bit rate (UBR). Of these, the ABR service has been specifically designed for efficient handling of data traffic.

One of the challenges in designing ATM traffic management was to maintain the QoS for various classes while attempting to make maximal use of network resources. This is what distinguishes traffic management from the "congestion control" problem of the past. Congestion control deals only with the problem of reducing load during overload. Traffic management deals not only with load reduction under overload or load increase during underload but more importantly it tries to ensure that the QoS guarantees are met in spite of varying load conditions. Thus, traffic management is required even if the network is underloaded.

---
[1]IEEE Communications Magazine, November 1996. Available through http://www.cis.ohio-state.edu/~jain/papers.html

This paper provides insights in to the development of ABR traffic management and explains reasons behind various decisions. The basic model used is introduced in the next section.

## 2 ABR Rate-based Traffic Management Model

ABR mechanisms allow the network to divide the available bandwidth fairly and efficiently among the active traffic sources. In the ABR traffic management framework, the *source end systems* limit their data transmission to rates allowed by the network. The network consists of *switches* which use their current load information to calculate the allowable rates for the sources. These rates are sent to the sources as feedback via *resource management (RM)* cells. RM cells are generated by the sources and travel along the data path to the *destination end systems*. The destinations simply return the RM cells to the sources. The components of the ABR traffic management framework are shown in Figure 1. In this tutorial, we explain the source and destination end-system behaviors and their implications on ABR traffic management.

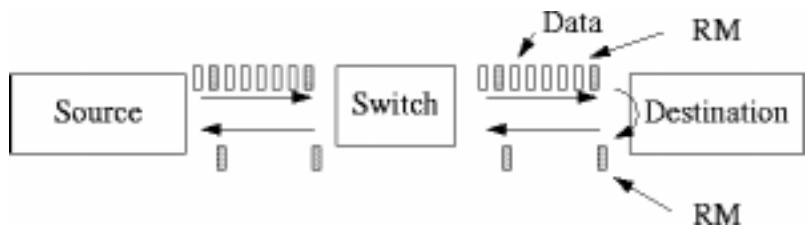

Figure 1: ABR Traffic Management Model: Source, Switch, Destination and Resource Management Cells

The ABR traffic management model is called a "rate-based end-to-end closed-loop" model. The model is called "rate-based" because the sources send data at a specified "rate." This is different from current packet networks (for example, TCP), where the control is "window based" and the sources limit their transmission to a particular number of packets. The ABR model is called "closed-loop" because there is a continuous feedback of control information between the network and the source. If more sources become active, the rate allocated to each source is reduced. The model used for CBR and VBR traffic, on the other hand, is "open-loop" in the sense that rates are negotiated at the beginning of the connection and do not change dynamically. Finally, the model is called "end-to-end" because the control cells travel from the source to the destination and back to the source. The alternative of "hop-by-hop" control in which each switch would give feedback to the previous switch was considered and not accepted due to its complexity. However, one can achieve the hop-by-hop control in TM4.0 using the virtual source/virtual destination (VS/VD) feature discussed later in this section.

When there is a steady flow of RM cells in the forward and reverse directions, there is a steady flow of feedback from the network. In this state, the ABR control loop has been established and the source rates are primarily controlled by the network feedback (closed-loop control). However, until the first RM cell returns, the source rate is controlled by the negotiated parameters, which may or may not relate to the current load on the network. The virtual circuit (VC) is said to be following an "open-loop" control during this phase. This phase normally lasts for one round-trip time (RTT). As we explain later, ABR sources are required to return to the open-loop control after long idle intervals. Traffic sources that have active periods (bursts) when data is transmitted at the allocated rate and idle periods when no data is transmitted are called "bursty sources" Open-loop control has a significant influence on the performance of bursty traffic particularly if it consists of bursts separated by long idle intervals.

There are three ways for switches to give feedback to the sources. First, each cell header contains a bit called Explicit Forward Congestion Indication (EFCI), which can be set by a congested switch. This mechanism is a modification of the DECbit scheme [12]. Such switches are called "binary" or "EFCI" switches. Second, RM cells have two bits in their payload, called the Congestion Indication (CI) bit and the No Increase (NI) bit, that can be set by congested switches. Switches that use only this mechanism are called relative rate marking switches. Third, the RM cells also have another field in their payload called explicit rate (ER) that can be reduced by congested switches to any desired value. Such switches are called explicit rate switches.

Explicit rate switches normally wait for the arrival of an RM cell to give feedback to a source. However, under extreme congestion, they are allowed to generate an RM cell and send it immediately to the source. This optional mechanism is called backward explicit congestion notification (BECN).

Switches can use the virtual source/virtual destination (VS/VD) feature to segment the ABR control loop into smaller loops. In a VS/VD network, the switches additionally behave both as a (virtual) destination end system and as a (virtual) source end system. As a destination end system, it turns around the RM cells to the sources from one segment. As a source end system, it generates RM cells for the next segment. This feature can allow feedback from nearby switches to reach sources faster, and allow hop-by-hop control as discussed earlier.

# 3  ABR Parameters

At the time of connection setup, ABR sources negotiate several operating parameters with the network. The first among these is the peak cell rate (PCR). This is the maximum rate at which the source will be allowed to transmit on this virtual circuit (VC). The source can also request a minimum cell rate (MCR) which is the guaranteed minimum rate. The network has to reserve this bandwidth for the VC. During the data transmission stage, the rate at which a source is allowed to send at any particular instant is called the allowed cell rate (ACR). The ACR is dynamically changed between MCR and PCR. At the beginning of the connection, and after long idle intervals, ACR is set to initial cell rate (ICR).

Table 1: List of ABR Parameters

| Label | Expansion | Default Value |
|---|---|---|
| PCR | Peak Cell Rate | - |
| MCR | Minimum Cell Rate | 0 |
| ACR | Allowed Cell Rate | - |
| ICR | Initial Cell Rate | PCR |
| TCR | Tagged Cell Rate | 10 cells/s |
| Nrm | Number of cells between FRM cells | 32 |
| Mrm | Controls bandwidth allocation between FRM, BRM and data cells | 2 |
| Trm | Upper Bound on Inter-FRM Time | 100 ms |
| RIF | Rate Increase Factor | 1/16 |
| RDF | Rate Decrease Factor | 1/16 |
| ADTF | ACR Decrease Time Factor | 0.5 ms |
| TBE | Transient Buffer Exposure | 16,777,215 |
| CRM | Missing RM-cell Count | $\lceil$ TBE/Nrm $\rceil$ |
| CDF | Cutoff Decrease Factor | 1/16 |
| FRTT | Fixed Round-Trip Time | - |

During the development of the RM specification, all numerical values in the specification were replaced by mnemonics. For example, instead of saying "every 32nd cell should be an RM cell" the specification states "every Nrm*th* cell should be an RM cell." Here, Nrm is a parameter whose default value is 32. Some of the parameters are fixed while others are negotiated. This being a tutorial (and not a standard document), we have reverted back to the default values of these parameters. This makes it easier to understand. A complete list of parameters used in the ABR mechanism is presented in Table 1. The parameters are explained as they occur in our discussion.

# 4  In-Rate and Out-of-Rate RM Cells

Most resource management cells generated by the sources are counted as part of their network load in the sense that the total rate of data and RM cells should not exceed the ACR of the source. Such RM cells are called "in-rate" RM cells. Under exceptional circumstances, switches, destinations, or even sources can generate extra RM cells. These "out-of-rate" RM cells are not counted in the ACR of the source and are distinguished by having their cell loss priority (CLP) bit set, which means that the network will carry them only if there is plenty of bandwidth and can discard them if congested. The out-of-rate RM cells generated by the source and switch are limited to 10 RM cells per second per VC. One use of out-of-rate RM cells is for BECN from the switches. Another use is for a source, whose ACR has been set to zero by the network, to periodically sense the state of the network. Out-of-rate RM cells are also used by destinations of VCs whose reverse direction ACR is either zero or not sufficient to return all RM cells received in the forward direction.

Note that in-rate and out-of-rate distinction applies only to RM cells. All data cells in ABR should have CLP set to 0 and must always be within the rate allowed by the network.

## 5 Forward and Backward RM cells

Resource Management cells traveling from the source to the destination are called Forward RM (FRM) cells. The destination turns around these RM cells and sends them back to the source on the same VC. Such RM cells traveling from the destination to the source are called Backward RM (BRM) cells. Forward and backward RM cells are illustrated in Figure 2. Note that when there is bi-directional traffic, there are FRMs and BRMs in both directions on the Virtual Channel (VC). A bit in the RM cell payload indicates whether it is an FRM or BRM. This direction bit (DIR) is changed from 0 to 1 by the destination.

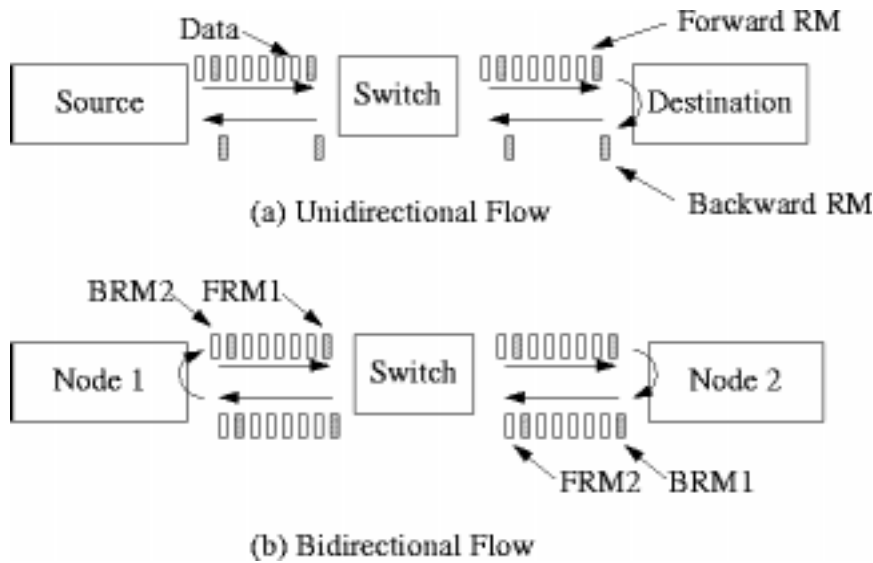

Figure 2: Forward and Backward Resource Management Cells (FRMs and BRMs)

## 6 RM Cell Format

The complete format of the RM cells is shown in figure 3. Every RM cell has the regular ATM header of five bytes. The payload type indicator (PTI) field is set to 110 (binary) to indicate that the cell is an RM cell. The protocol id field, which is one byte long, is set to one for ABR connections. The direction (DIR) bit distinguishes forward and backward RM cells. The backward notification (BN) bit is set only in switch generated BECN cells. The congestion indication (CI) bit is used by relative rate marking switches. It may also be used by explicit rate switches under extreme congestion as discussed later. The no increase (NI) bit is another bit available to explicit rate switches to indicate moderate congestion. The request/acknowledge, queue length, and sequence number fields of the RM cells are for compatibility with the ITU-T recommendation I.371 and are not used by the ATM Forum.

The Current Cell Rate (CCR) field is used by the source to indicate to the network its current rate. Some switches may use the CCR field to determine a VC's next allocation while others may measure the VC's rate and not trust CCR. The minimum cell rate (MCR) field is redundant in the sense that like PCR, ICR, and other parameters it does not change during the life of a connection. However, its presence in the RM cells reduces number of lookups required in the switch.

The Explicit Rate (ER) field, the CI and the NI fields are used by the network to give feedback to the sources. The ER field indicates the maximum rate allowed to the source. When there are multiple switches along the path, the feedback given by the most congested link is the one that reaches the source.

Data cells also have an Explicit Forward Congestion Indication (EFCI) bit in their headers, which may be set by the network when it experiences congestion. The destination saves the EFCI state of every data cell. If the EFCI state is set when it turns around an RM cell, it uses the CI bit to give (a single bit) feedback to the source. When the source receives the RM cell from the network, it adjusts its ACR using the ER, CI, NI values, and source parameters.

| Field | Size | |
|---|---|---|
| ATM Header | 5 Bytes | |
| Protocol ID | 1 Byte | 1 = ABR |
| Direction | 1 bit | 0 = Forward |
| Backward Notification | 1 bit | 1 = Switch/dest generated |
| Congestion Indication | 1 bit | 1 = High Congestion |
| No Increase | 1 bit | 1 = Mild congestion |
| Request/Acknowledge* | 1 bit | |
| Reserved | 3 bits | |
| Explicit Rate | 2 Bytes | |
| Current Cell Rate | 2 Bytes | |
| Minimum Cell Rate | 2 Bytes | |
| Queue Length* | 4 Bytes | |
| Sequence Number* | 4 Bytes | |
| Reserved | 30.75 Bytes | |
| CRC-10 | 10 bits | |

Figure 3: Resource Management (RM) Cell Fields

All rates (for example, ER, CCR, and MCR) in the RM cell are represented using a special 16-bit floating point format, which allows a maximum value of 4,290,772,992 cells per second (1.8 terabits per second). During connection setup, however, rate parameters are negotiated using an 24-bit integer format, which limits their maximum value to 16,777,215 cells per second or 7.1 Gbps.

# 7 Source End System Rules

TM4.0 specifies 13 rules that the sources have to follow. This section discusses each rule and traces the development and implications of certain important rules. In some cases the precise statement of the rule is important. Hence, the source and destination rules are quoted from the TM specification [1] in appendix 9. A list of abbreviations and their expansions is provided in appendix 9.

- **Source Rule 1**: Sources should always transmit at a rate equal to or below their computed ACR. The ACR cannot exceed PCR and need not go below MCR. Mathematically,
$$\text{MCR} \leq \text{ACR} \leq \text{PCR}$$
$$\text{Source Rate} \leq \text{ACR}$$

- **Source Rule 2**: At the beginning of a connection, sources start at ICR. The first cell is always an in-rate forward RM cell. This ensures that the network feedback will be received as soon as possible.

- **Source Rule 3**: At any instant, sources have three kinds of cells to send: data cells, forward RM cells, and backward RM cells (corresponding to the reverse flow). The relative priority of these three kinds of cells is different at different transmission opportunities.

  First, the sources are required to send an FRM after every 31 cells. However, if the source rate is low, the time between RM cells will be large and network feedback will be delayed. To overcome this problem, a source is supposed to send an FRM cell if more than 100 ms has elapsed since the last FRM. This introduces another problem for low rate sources. In some cases, at every transmission opportunity the source may find that it has exceeded 100 ms and needs to send an FRM cell. In this case, no data cells will be transmitted. To overcome this problem, an additional condition was added that there must be at least two other cells between FRMs.

  An example of the operation of the above condition is shown in the figure 4. The figure assumes a unidirectional VC (i.e., there are no BRMs to be turned around). The figure has three parts. The first part of the figure shows that, when the source rate is 500 cells/s, every 32nd cell is an FRM cell. The time to send 32 cells is always smaller than 100 ms. In the second part of the figure, the source rate is 50 cells/s. Hence 32 cells takes 640 ms to be transmitted. Therefore, after 100 ms, an FRM is scheduled in the next transmission opportunity (or slot). The third part of the figure shows the scenario when the source rate is 5 cells/s. The inter-cell time itself is 200 ms. In this case, an FRM is scheduled every three slots, i.e., the inter-FRM time is 600 ms.

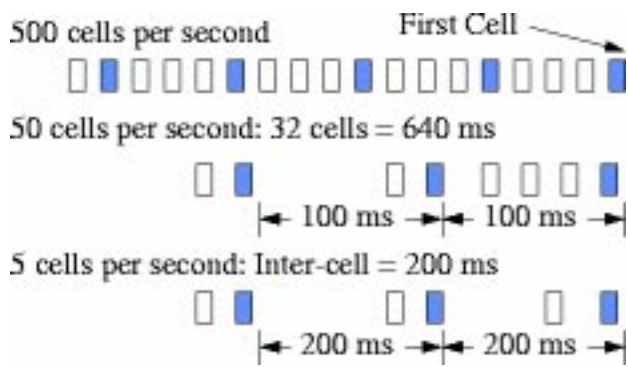

Figure 4: Frequency of forward RM cells.

Second, a waiting BRM has priority over waiting data, given that no BRM has been sent since the last FRM. Of course, if there are no data cells to send, waiting BRMs may be sent.

Third, data cells have priority in the remaining slots.

The second and third part of the this rule ensure that BRMs are not unnecessarily delayed and that all available bandwidth is not used up by the RM cells.

Figure 5 illustrates the scheduling of FRMs, BRMs and data cells. In the first slot, an FRM is scheduled. In the next slot, assuming that a turned around BRM is awaiting transmission, a BRM is scheduled. In the remaining slots data is scheduled. If the rate is low, more FRMs and BRMs may be scheduled.

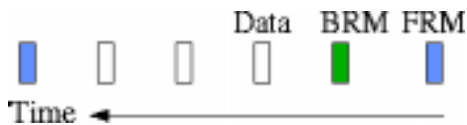

Figure 5: Scheduling of forward RM, backward RM, and data cells.

- **Source Rule 4**: All RM cells sent in accordance with rules 1-3 are in-rate RM cells and have their cell loss priority (CLP) bit set to 0. Additional RM cells may be sent out-of-rate and should have their CLP bit set to 1. For example, consider the third unidirectional flow of Figure 4. It has an ACR of 5 cells/s. It is allowed to send only one in-rate RM cell every 400 ms. If necessary, it can send a limited number of out-of-rate RM cells with CLP set to 1.

The frequency of FRM is determined by parameters Nrm, Trm, and Mrm, whose default values are 32, 100 ms, and 2, respectively. During the debate on credit vs rate based alternatives for traffic management [5], the rate based group selected a default value of 32 for Nrm. This ensured that the control overhead was equivalent to that of credit based alternative which claimed an overhead of approximately 6%. During normal operation 1/32th or 3% of all cells are FRM cells. Similarly, another 3% of cells are BRM cells resulting in a total overhead of 6%.

In practice, the choice of Nrm affects the responsiveness of the control and the computational overhead at the end systems and switches. For a connection running at 155 Mbps, the inter-RM cell time is 86.4 $\mu$s while it is 8.60 ms for the same connection running at 1.55 Mbps. The inter-RM interval determines the responsiveness of the system. While most end-systems and switches will do ABR computations in hardware, it has been shown that it is possible to do them in software on a Pentium$^{TM}$ system provided Nrm is set to 192 or higher on a 155 Mbps link.

- **Source Rule 5**: The rate allowed to a source is valid only for approximately 500 ms. If a source does not transmit any RM cells for this duration, it cannot use its previously allocated ACR particularly if the ACR is high. The source should re-sense the network state by sending an RM cell and decreasing its rate to the initial cell rate (ICR) negotiated at connection setup. If a source's ACR is already below ICR, it should stay at that lower value (and not increase it to ICR).

  The timeout interval is set by the ACR Decrease Time Factor (ADTF). This parameter can be negotiated with the network at connection setup. Its default value is 500 ms.

  This simple rule was the cause of a big debate at the Forum. It is intended to solve the problem of *ACR retention*. If a source sends an RM cell when the network is not heavily loaded, the source may be granted a very high rate. The source can then retain that rate and use it when the network is highly loaded. In fact, a source may set up several VCs and use them to get an unfair advantage. To solve this problem, several so called *use it or lose it* (UILI) solutions were proposed. Some of them relied on actions at the source while others relied on actions at the switch. The source based solutions required sources to monitor their own rates and reduce ACR slowly if was too high compared to the rate used.

  UILI alternatives were analyzed and debated for months because they have a significant impact on the performance of bursty traffic that forms the bulk of data traffic.

  The ATM Forum chose to standardize a very simple UILI policy at the source. This policy provided a simple timeout method (using ADTF as the timeout value) which reduces ACR to ICR when the timeout expires. Vendors are free to implement additional proprietary restraints at the source or at the switch. A few examples of such possibilities are listed in the Informative Appendix I.8 of the specification [1]. See also [4], [2], [11].

- **Source Rule 6**: If a network link becomes broken or becomes highly congested, the RM cells may get blocked in a queue and the source may not receive the feedback. To protect the network from continuous in-flow of traffic under such circumstances, the sources are required to reduce their rate if the network feedback is not received in a timely manner.

Normally under steady state, sources should receive one BRM for every FRM sent. Under congestion, BRM cells may be delayed. If a source has sent CRM FRM cells and has not received any BRM, it should suspect network congestion and reduce its rate by a factor of CDF. Here, CRM (missing RM cell count) and CDF (cutoff decrease factor) are parameters negotiated at the time of connection setup. BECN cells generated by switches (and identified by BN=1) are not counted as BRM.

When rule 6 triggers once, the condition is satisfied for all successive FRM cells until a BRM is received. Thus, this rule results in a fast exponential decrease of ACR.

An important side effect of this rule is that unless CRM is set high, the rule could trigger unnecessarily on a long delay path. CRM is computed from another parameter called transient buffer exposure (TBE) which is negotiated at connection setup. TBE determines the maximum number of cells that may suddenly appear at the switch during the first round trip before the closed-loop phase of the control takes effect. During this time, the source will have sent TBE/Nrm RM cells. Hence,

$$\text{CRM} = \lceil \frac{\text{TBE}}{\text{Nrm}} \rceil$$

The fixed part of the round-trip time (FRTT) is computed during connection setup. This is the minimum delay along the path and does not include any queueing delay. During this time, a source may send as many as ICR × FRTT cells into the network. Since this number is negotiated separately as TBE, the following relationship exists between ICR and TBE:

$$ICR \times FRTT \leq TBE$$

or

$$ICR \leq TBE/FRTT$$

The sources are required to use the ICR value computed above if it is less than the ICR negotiated with the network. In other words:

ICR used by the source = Min{ICR negotiated with the network, TBE/FRTT}

In negotiating TBE, the switches have to consider their buffer availability. As the name indicates, the switch may be suddenly exposed to TBE cells during the first round trip (and also after long idle periods). For small buffers, TBE should be small and vice versa. On the other hand, TBE should also be large enough to prevent unnecessary triggering of rule 6 on long delay paths.

It has been incorrectly believed that cell loss could be *avoided* by simply negotiating a TBE value below the number of available buffers in the switches. Jain et al. [10] showed that it is possible to construct workloads where queue sizes could be unreasonably high even when TBE is very small. For example, if the FRM input rate is $x$ times the BRM output rate (see Figure 6), where $x$ is less than CRM, rule 6 will not trigger but the queues in the network will keep building up at the rate of $(x - 1) \times$ ACR leading to large queues. The only reliable way to protect a switch from large queues is to build it in the switch allocation algorithm. The ERICA+ algorithm [7] is an example of one such algorithm.

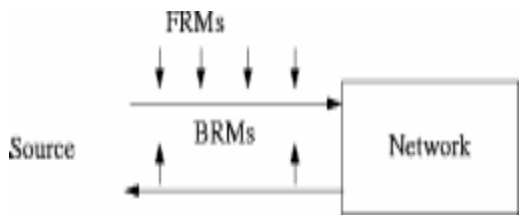

Figure 6: Source Rule 6 does not trigger if BRM flow is maintained

Observe that the FRTT parameter which is the sum of fixed delays on the path is used in the formula for ICR. During the development of this rule, an estimate of round trip time (RTT), including the fixed and variable delays was being used instead of FRTT in the ICR calculation. We argued that RTT estimated at connection setup is a random quantity bearing little relation to the round trip delays during actual operation [9]. Such parameter setting could trigger source Rule 6 unnecessarily and degrade performance. Hence, the Forum decided to use FRTT parameter instead of RTT.

Note that it is possible to disable source Rule 6, by setting CDF to zero.

- **Source Rule 7**: When sending an FRM, the sources should indicate their current ACR in the CCR field of the RM cells.

- **Source Rules 8 and 9**: Source Rule 8 and 9 describe how the source should react to network feedback. The feedback consists of explicit rate (ER), congestion indication bit (CI), and no-increase bit (NI). Normally, a source could simply change its ACR to the new ER value but this could cause a few problems as discussed next.

  First, if the new ER is very high compared to current ACR, switching to the new ER will cause sudden queues in the network. Therefore, the amount of increase is limited. The rate increase factor (RIF) parameter determines the maximum allowed increase in any one step. The source cannot increase its ACR by more than RIF $\times$ PCR.

  Second, if there are any EFCI switches in the path, they do not change the ER field. Instead, they set EFCI bits in the cell headers. The destination monitors these bits and returns the last seen EFCI bit in the CI field of a BRM. A CI of 1 means that the network is congested and that the source should reduce its rate. The decrease is determined by rate decrease factor (RDF) parameter. Unlike the increase, which is additive, the decrease is multiplicative in the sense that

$$\text{ACR} \leftarrow \text{ACR}(1 - \text{RDF})$$

It has been shown that additive increase and multiplicative decrease is sufficient to achieve fairness [6]. Other combinations such as additive increase with additive decrease, multiplicative increase with multiplicative decrease, and multiplicative increase with additive increase are unfair.

The no-increase (NI) bit was introduced to handle mild congestion cases. In such cases, a switch could specify an ER, but instruct that, if ACR is already below the specified ER, the source should not increase the rate. The actions corresponding to the various values of CI and NI bits are as follows:

| NI | CI | Action |
|----|----|--------|
| 0  | 0  | ACR $\leftarrow$ Min(ER, ACR + RIF $\times$ PCR, PCR) |
| 0  | 1  | ACR $\leftarrow$ Min(ER, ACR $-$ ACR $\times$ RDF) |
| 1  | 0  | ACR $\leftarrow$ Min(ER, ACR) |
| 1  | 1  | ACR $\leftarrow$ Min(ER, ACR $-$ ACR $\times$ RDF) |

$$ACR \leftarrow Max(ACR, MCR)$$

If there are no EFCI switches in a network, setting RIF to 1 allows ACRs to increase as fast as the network directs it. This allows the available bandwidth to be used quickly. For EFCI networks, or a combination of ER and EFCI networks, RIF should be set conservatively to avoid unnecessary oscillations.

Once the ACR is updated, the subsequent cells sent from the source conform to the new ACR value. However, if the earlier ACR was very low, it is possible that the very next cell is scheduled a long time in the future. In such a situation, it is advantageous to "reschedule" the next cell, so that the source can take advantage of the high ACR allocation immediately [8].

- **Source Rule 10**: Sources should initialize various fields of FRM cells as follows. For virtual path connections (VPCs), the virtual circuit id (VCI) is set to 6. For virtual channel connections (VCCs), the VCI of the connection is used. In either case, the protocol type id (PTI) in the ATM cell header is set to 6 (110). The protocol id field in the payload of the RM cell is set to 1. The direction bit should be set to 0 (forward). The backward notification (BN) bits should be set to 0 (source generated). Explicit rate field is initialized to the maximum rate below PCR that the source can support. Current cell rate is set to current ACR. Minimum cell rate is set to the value negotiated at connection setup. Queue length, sequence number, and request/acknowledge fields are set in accordance with ITU-T recommendation I.371 or to zero. All reserved octets are set to 6A (hex) or 01101010 (binary). This value is specified in ITU-T recommendation I.610 (whose number coincidently is also 6-A in hex). Other reserved bits are set to 0. Note that the sources are allowed to set ER and NI fields to indicate their own congestion.

- **Source Rule 11**: The out-of-rate FRM cells generated by sources are limited to to a rate below the "tagged cell rate (TCR)" parameter, which has a default value of 10 cells per second.

- **Source Rule 12**: The EFCI bit must be reset on every data cell sent. The alternative of congested sources being allowed to set EFCI bit was considered but rejected due to insufficient analysis.

- **Source Rule 13**: Sources can optionally implement additional Use-It-or-Lose-It (UILI) policies (see discussion of source Rule 5).

# 8 Destination End System Rules

- **Destination Rule 1**: Destinations should monitor the EFCI bits on the incoming cells and store the value last seen on a data cell.

- **Destination Rule 2**: Destinations are required to turn around the forward RM cells with minimal modifications as follows: the DIR bit is set to "backward" to indicate that the cell is a backward RM-cell; the BN bit is set to zero to indicate that the cell was not generated by a switch; the CCR and MCR fields should not be changed.

  If the last cell has EFCI bit set, the CI bit in the next BRM is set and the stored EFCI state is cleared.

  If the destination has internal congestion, it may reduce the ER or set the CI or NI bits just like a switch. Observe that this rule is used in the VS/VD configuration where the virtual destination is bottlenecked by the allowed rate in the next segment. In any case, the ER is never increased.

- **Destination Rules 3-4**: The destination should turn around the RM cells as fast as possible. However, an RM cell may be delayed if the reverse ACR is low. In such cases destination rules 3 and 4 specify that old out-of-date information can be discarded. The destinations are allowed a number of options to do this. The implications of various options of destination Rule 3 are discussed in the Informative Appendix I.7 of the TM specification [1]. Briefly, the recommendations attempt to ensure the flow of feedback to the sources for a wide range of values of ACR of the reverse direction VC. If the reverse direction ACR is non-zero, then a backward RM cell will be scheduled for in-rate transmission. Transmitting backward RM cells out-of-rate ensures that the feedback is sent regularly even if the reverse ACR is low or zero (for example, in unidirectional VCs).

  Note that there is no specified limit on the rate of such "turned around" out-of-rate RM cells. However, the CLP bit is set to 1 in the out-of-rate cells, which allows them to be selectively dropped by the switch if congestion is experienced.

- **Destination Rule 5**: Sometimes a destination may be too congested and may want the source to reduce its rate immediately without having to wait for the next RM cell. Therefore, like a switch, the destinations are allowed to generate BECN RM cells. Also, as in the case of switch generated BECNs, these cells may not ask a source to increase its rate (CI bit is set). These BECN cells are limited to 10 cells/s and their CLP bits are set (i.e., they are sent out-of-rate).

- **Destination Rule 6**: An out-of-rate FRM cell may be turned around either in-rate (with CLP=0) or out-of-rate (with CLP=1).

# 9 Summary

We have presented the source and destination rules and parameters of the ABR traffic management model. The history and reasons behind these rules were explained. Like any other standard, these rules reflect a compromise between several differing views. We have tried to reflect these differing views. Implementation experience in the next few years will help us further understand the importance of various rules and parameters.

---

[2] All our papers and ATM Forum contributions are available through the www site: http://www.cis.ohio-state.edu/~jain/

# Appendix A: Source and Destination Behavior

This appendix provides the precise source and destination behavior verbatim from the TM specification [1]. All table, section, and other references in this appendix refer to those in the TM specification.

## 5.10.4 Source Behavior

The following items define the source behavior for CLP=0 and CLP=1 cell streams of a connection. By convention, the CLP=0 stream is referred to as in-rate, and the CLP=1 stream is referred to as out-of-rate. Data cells shall not be sent with CLP=1.

1. The value of ACR shall never exceed PCR, nor shall it ever be less than MCR. The source shall never send in-rate cells at a rate exceeding ACR. The source may always send in-rate cells at a rate less than or equal to ACR.

2. Before a source sends the first cell after connection setup, it shall set ACR to at most ICR. The first in-rate cell shall be a forward RM-cell.

3. After the first in-rate forward RM-cell, in-rate cells shall be sent in the following order:

   **a)** The next in-rate cell shall be a forward RM cell if and only if, since the last in-rate forward RM-cell was sent, either:

      **i)** at least Mrm in-rate cells have been sent and at least Trm time has elapsed, or

      **ii)** Nrm -1 in-rate cells have been sent.

   **b)** The next in-rate cell shall be a backward RM-cell if condition (a) above is not met, if a backward RM cell is waiting for transmission, and if either:

      **i)** no in-rate backward RM-cell has been sent since the last in-rate forward RM-cell, or

      **ii)** no data cell is waiting for transmission.

   **c)** The next in-rate cell sent shall be a data cell if neither condition (a) nor condition (b) is met, and if a data cell is waiting for transmission.

4. Cells sent in accordance with source behaviors #1,#2, and #3 shall have CLP=0.

5. Before sending a forward in-rate RM cell, if ACR > ICR and the time T that has elapsed since the last in-rate forward RM-cell was sent is greater than ADTF, then ACR shall be reduced to ICR.

6. Before sending an in-rate forward RM cell, and following behavior #5 above, if at least CRM in-rate forward RM-cells have been sent since the last backward RM-cell with BN=0 was received, then ACR shall be reduced by at least ACR × CDF, unless that reduction would result in a rate below MCR, in which case ACR shall be set to MCR.

7. After following behaviors #5 and #6 above, the ACR value shall be placed in the CCR field of the outgoing forward RM-cell, but only in-rate cells sent after the outgoing forward RM-cell need to follow the new rate.

8. When a backward RM-cell (in-rate or out-of-rate) is received with CI=1, then ACR shall be reduced by at least ACR × RDF, unless that reduction would result in a rate below MCR, in which case ACR shall be set to MCR. If the backward RM-cell has both CI=0 and NI=0, then the ACR may be increased by no more than RIF × PCR, to a rate not greater than PCR. If the backward RM-cell has NI=1, the ACR shall not be increased.

9. When a backward RM-cell (in-rate or out-of-rate) is received, and after ACR is adjusted according to source behavior #8, ACR is set to at most the minimum of ACR as computed in source behavior #8, and the ER field, but no lower than MCR.

10. When generating a forward RM-cell, the source shall assign values to the various RM-cell fields as specified for source-generated cells in Table 5-4.

11. Forward RM-cells may be sent out-of-rate (i.e., not conforming to the current ACR). Out-of-rate forward RM-cells shall not be sent at a rate greater than TCR.

12. A source shall reset EFCI on every data cell it sends.

13. The source may implement a use-it-or-lose-it policy to reduce its ACR to a value which approximated the actual cell transmission rate. Use-it-or-lose-it policies are discussed in Appendix I.8.

**Notes:**

1. In-rate forward and backward RM-cells are included in the source rate allocated to a connection.

2. The source is responsible for handling congestion within its scheduler in a fair manner. This congestion occurs when the sum of the rates to be scheduled exceeds the output rate of the scheduler. The method for handling local congestion is implementation specific.

### 5.10.5 Destination Behavior

The following items define the destination behavior for CLP=0 and CLP=1 cell streams of a connection. By convention, the CLP=0 stream is referred to as in-rate, and the CLP=1 stream is referred to as out-of-rate.

1. When a data cell is received, its EFCI indicator is saved as the EFCI state of the connection.

2. On receiving a forward RM-cell, the destination shall turn around the cell to return to the source. The DIR bit in the RM-cell shall be changed from "forward" to "backward," BN shall be set to zero, and CCR, MCR, ER, CI, and NI fields in the RM-cell shall be unchanged except:

   a) If the saved EFCI state is set, then the destination shall set CI=1 in the RM cell, and the saved EFCI state shall be reset. It is preferred that this step is performed as close to the transmission time as possible;

   b) The destination (having internal congestion) may reduce ER to whatever rate it can support and/or set CI=1 or NI=1. A destination shall either set the QL and SN fields to zero, preserve these fields, or set them in accordance with ITU-T Recommendation I.371-draft. The octets defined in Table 5-4 as reserved may be set to 6A (hexadecimal) or left unchanged. The bits defined as reserved in Table 5-4 for octet 7 may be set to zero or left unchanged. The remaining fields shall be set in accordance with Section 5.10.3.1 (Note that this does not preclude looping fields back from the received RM cell).

3. If a forward RM-cell is received by the destination while another turned-around RM-cell (on the same connection) is scheduled for in-rate transmission:

   a) It is recommended that the contents of the old cell are overwritten by the contents of the new cell;

   b) It is recommended that the old cell (after possibly having been overwritten) shall be sent out-of-rate; alternatively the old cell may be discarded or remain scheduled for in-rate transmission;

   c) It is required that the new cell be scheduled for in-rate transmission.

4. Regardless of the alternatives chosen in destination behavior #3, the contents of the older cell shall not be transmitted after the contents of a newer cell have been transmitted.

5. A destination may generate a backward RM-cell without having received a forward RM-cell. The rate of the backward RM-cells (including both in-rate and out-of-rate) shall be limited to 10 cells/second, per connection. When a destination generated an RM-cell, it shall set either CI=1 or NI=1, shall set set BN=1, and shall set the direction to backward. The destination shall assign values to the various RM-cell fields as specified for destination generated cells in Table 5-4.

6. When a forward RM-cell with CLP=1 is turned around it may be sent in-rate (with CLP=0) or out-of-rate (with CLP=1)

**Notes**

1. "Turn around" designates a destination process of transmitting a backward RM-cell in response to having received a forward RM-cell.

2. It is recommended to turn around as many RM-cells as possible to minimize turn-around delay, first by using in-rate opportunities and then by using out-of-rate opportunities as available. Issues regarding turning RM-cells around are discussed in Appendix I.7.

# Appendix B: List of Abbreviations

| | |
|---|---|
| ABR | Available bit rate |
| ACR | Allowed cell rate |
| ADTF | ACR decrease time factor |
| ATM | Asynchronous transfer mode |
| BECN | Backward explicit congestion notification |
| BN | Backward notification |
| BRM | Backward resource management cell |
| CBR | Constant bit rate |
| CCR | Current cell rate |
| CDF | Cutoff decrease factor |
| CI | Congestion indication |
| CLP | Cell loss priority |
| CRM | Missing RM cell count |
| DIR | Direction |
| EFCI | Explicit forward congestion indication |
| ER | Explicit rate |
| ERICA | Explicit rate indication for congestion avoidance |
| FRM | Forward resource management cell |
| FRTT | Fixed round-trip time |
| ICR | Initial Cell Rate |
| ITU | International Telecommunication Union |
| MCR | Minimum cell rate |
| Mrm | Minimum number of cells between FRM cells |
| NI | No increase |
| Nrm | Number of cells between FRM cells |
| PCR | Peak cell rate |
| PTI | Protocol type identifier |
| QL | Queue length |
| RDF | Rate decrease factor |
| RIF | Rate increase factor |
| RM | Resource management cell |
| RTT | Round-trip time |
| TBE | Transient buffer exposure |
| TCP | Transport Control Protocol |
| TCR | Tagged cell rate |
| TM | Traffic management |
| Trm | Upper bound on inter-FRM time |
| UBR | Unspecified bit rate |
| UILI | Use it or lose it |
| VBR | Variable bit rate |
| VC | Virtual channel |
| VCC | Virtual channel connection |
| VCI | Virtual channel identifier |
| VD | Virtual destination |
| VPC | Virtual path connection |
| VS | Virtual source |

**Raj Jain** is a Professor of Computer Information Sciences at The Ohio State University in Columbus, Ohio. Prior to joining the university in April 1994, he was a Senior Consulting Engineer at Digital Equipment Corporation, Littleton, Massachusetts, where he was involved in design and analysis of many computer systems and networks including VAX Clusters, Ethernet, DECnet, OSI, FDDI, and ATM networks. Currently he is very active in the Traffic Management working group of ATM Forum and has influenced its direction considerably. Raj Jain received a Ph.D. degree in Computer Science from Harvard University in 1978. He taught at the Massachusetts Institute of Technology in 1984, 1985, and 1987. He is the author of two books: "The Art of Computer Systems Performance Analysis," and "FDDI Handbook: High-Speed Networking with Fiber and Other Media." He is an IEEE Fellow, an ACM Fellow and is on the editorial boards of several journals. He has several patents and over 35 publications.

**Shivkumar Kalyanaraman** is a doctoral candidate in Computer Sciences at the Ohio state University. He received his B.Tech degree from the Indian Institute of Technology, Madras, India in July 1993. He received his M.S. degree from the Ohio State University in 1994. His research interests include broadband networks, transport protocols, congestion control, distributed systems, and performance analysis. He is a co-inventor of two patents, and has co-authored several papers and ATM forum contributions in the field of ATM congestion control. He is a student member of IEEE-CS and ACM.

**Sonia Fahmy** received her BS degree from the American University in Cairo, Egypt, in June 1992, and her MS degree from the Ohio State University in March 1996, both in Computer Science. She is currently a PhD student at the Ohio State University. Her main research interests are in the areas of broadband networks, congestion control, performance analysis, distributed computing and programming languages. She co-authored several papers and ATM Forum contributions. She is a student member of the ACM and the IEEE-CS.

**Rohit Goyal** is a PhD. student with the Department of Computer and Information Science at the Ohio State University, Columbus. He received his BS in Computer Science from Denison University, Granville. His work has been published in several ATM forum contributions and Technical Reports. His other interests include Distributed Systems, Artificial Intelligence, and Performance Analysis. Rohit is a member of Phi Beta Kappa, Who's who among students in American Colleges, Sigma Xi, Pi Mu Epsilon, Sigma Pi Sigma and the Phi Society.

**Seong-Cheol Kim** received his Ph.D from Polytechnic University in 1995 in electrical engineering. Since 1995 he has been with Samsung Electronics as a principal engineer. His research interests include traffic control, congestion control, and multimedia and ATM communications.